\documentclass{iopart}

\usepackage{graphicx} 
\usepackage{float}
\usepackage{color}
\usepackage{multirow}
\usepackage{dcolumn}  

\begin{document}

\title[Magnetic dichroism study on Mn$_{1.8}$Co$_{1.2}$Ga thin film]
      {Magnetic dichroism study on Mn$_{1.8}$Co$_{1.2}$Ga thin film using a
       combination of X-ray absorption and photoemission spectroscopy}

\author{Siham Ouardi\textsuperscript{1},
        Gerhard H. Fecher\textsuperscript{1},
        Takahide Kubota\textsuperscript{2},
        Shigemi Mizukami\textsuperscript{2},
        Eiji Ikenaga\textsuperscript{3},
        Tetsuya Nakamura\textsuperscript{3},
        Claudia Felser\textsuperscript{1}}
\address{\textsuperscript{1}Max Planck Institute for Chemical Physics of Solids, 01187 Dresden, Germany}
\address{\textsuperscript{2}WPI-Advanced Institute for Materials Research (WPI-AIMR), Tohoku University, Sendai 980-8577, Japan}
\address{\textsuperscript{3}Japan Synchrotron Radiation Research Institute, SPring-8, Sayo, Hyogo 679-5198, Japan}

\ead{fecher@cpfs.mpg.de}

\date{\today}
\begin{abstract}

Using circularly polarised radiation and a combination of bulk-sensitive hard 
X-ray photoelectron spectroscopy and X-ray-absorption spectroscopy (XAS) we 
studied the electronic and magnetic structure of epitaxial 
Mn$_{1.8}$Co$_{1.2}$Ga thin films. Spin resolved Bloch spectral functions, 
density of states as well as charge and magnetisation densities were 
investigated by a first-principles analysis of full potential, fully 
relativistic Korringa--Kohn--Rostoker calculations of the electronic structure. 
The valence states were experimentally investigated by using linear dichroism in 
the angular distribution and comparing the results to spin-resolved densities of 
states. The linear dichroism in the valence band enabled a symmetry analysis of 
the contributing states. The spectra were in good agreement with the theoretical 
partial density of states. The element-specific, spin-resolved, unoccupied 
densities of states  for Co and Mn were analysed by using XAS and X-ray magnetic 
circular dichroism (XMCD) at the $L_{3,2}$ edges. The spectra were influenced by 
strong correlation effects. XMCD was used to extract the site resolved magnetic 
moments. The experimental values of $m_{\rm Mn}=0.7\:\mu_B$ and
$m_{\rm Co}=1.05\:\mu_B$ agree very well with the calculated magnetic moments. Magnetic 
circular dichroism in angle-resolved photoelectron spectroscopy at the Mn and Co 
$2p$ core level exhibited a pronounced magnetic dichroism and confirmed the 
localised character of the Mn $d$ valence states.

\end{abstract}



\submitto{\JPD}
\section{Introduction}

Beyond conventional perpendicular magnetic anisotropy (PMA) materials, 
tetragonal Mn-based Heusler alloys have recently opened up a new way to develop 
materials with high anisotropy~\cite{MWS10,MKW12}. Mn$_3$Ga has been proposed as 
a compensated ferrimagnet having a high spin polarisation of 88\%, hard magnetic 
properties, and a high Curie temperature $T_c$ (730~K)~\cite{BFW07,WBF08}. 
Epitaxial growth of Mn--Ga thin films with high PMA~\cite{WMW09,MKW12} and 
multilayer devices with high tunnelling magnetoresistance 
(TMR)~\cite{KAM11,KMM12,MWS11,MAS14} have been realised.

Recently, it was shown that partial substitution of Mn by Co in Mn$_{3-x}$Co$_x$Ga 
changes the crystal structure from tetragonal to cubic when the Co 
content $x$ is increased, leading to a reduction of the saturation 
magnetisation~\cite{AWF11,WCG12,OKF12}. Kubota {\it et al.}~\cite{KOM13} 
investigated the composition dependence of the structural, magnetic, and 
transport properties of epitaxially grown Mn--Co--Ga films. The resistivity of 
the Mn--Co--Ga films was larger than that of pure Mn--Ga films. The Co 
substitution results in a reduced electron mobility because of the presence of 
localised electron states around the Fermi energy $\epsilon_F$ (intrinsic 
factor). Investigation of the TMR in  Mn--Co--Ga/MgO/CoFeB magnetic tunnel 
junctions (MTJs) revealed lower TMR values ($\leq 11$\%) compared to Mn--Ga 
MTJs~\cite{KMM14}. For other Heusler-based MTJ systems, the Mn content strongly 
influences the TMR ratio, with a record TMR ratio of 1995\% being obtained for 
off-stoichiometric Co$_2$MnSi Heusler-based MTJs \cite{LHT12}. It was predicted 
by first-principle calculations that spin-transport in the Mn--Ga system may be 
improved by chemical disorder introduced by a partial substitution of Mn by a 
$3d$ element~\cite{CKF12}. Realising this idea requires a detailed investigation 
of the electronic and magnetic structure of the Mn--Co--Ga system.

Hard X-ray photoemission spectroscopy (HAXPES) is a powerful tool for 
investigation of the chemical states and electronic structure of various 
materials~\cite{Kob05,Kob09} for example 
multilayers~\cite{FBO07,FBG08,OBG09,VOK14}, strongly correlated 
oxides~\cite{KKO14}, or magnetic materials especially in combination with 
variable photon polarisation~\cite{SYH10,KFS11,OFK11,SHI13}. X-ray magnetic 
circular dichroism (XMCD) in X-ray absorption spectroscopy (XAS) is an efficient 
method for studying the element-specific electronic structure of buried 
layers~\cite{TKL06} using hard ($K$ edges of $3d$ materials)~\cite{SWW87} as 
well as soft X-rays ($L$ edges of $3d$ materials)~\cite{CSM90} for excitation. 
The $L$-edge absorption spectra for left and right circularly polarised X-rays 
reflect the spin-resolved partial density of states (PDOS).

The present study reports on a detailed investigation of the electronic and 
magnetic structure of epitaxial Mn$_{1.8}$Co$_{1.2}$Ga thin films. Spin-resolved 
densities of states are calculated based on a first-principles analysis with 
fully relativistic Korringa--Kohn--Rostoker calculations and the results are 
compared to experiment. Linearly polarised radiation in combination with bulk-sensitive
HAXPES is used to study the symmetry of the valence states. Magnetic 
circular dichroism (MCD) of the core states is used to explore the film's 
magnetic properties. Element-specific magnetic moments are investigated by 
circular dichroism in XAS.

\section{Experimental details}
\label{sec:expdetail}

Epitaxial, 30-nm-thick thin films with nominal compositions of 
Mn$_{1.8}$Co$_{1.2}$Ga were grown on a MgO(001) single crystalline substrate 
using an ultrahigh vacuum magnetron sputtering system. A Mn--Ga target and an 
elemental Co target were used for co-deposition. Thin MgO and Al layers were 
deposited on top of the films to prevent their oxidation. The samples had the 
following stacking order: \\ 
MgO(100) substrate~/ Mn$_{1.8}$Co$_{1.2}$Ga(30 nm)~/ MgO(2 nm)~/ Al(2 nm). \\
Additional details about the film growth were previously reported in 
References~\cite{OKF12,KOM13}. The composition of the film --~as determined by 
inductively coupled plasma mass spectroscopy (ICP-MS)~-- was 
Mn$\,$:$\,$Co$\,$:$\,$Ga = 1.8$\,$:$\,$1.2$\,$:$\,$1. The nominally cubic 
structure has a very small tetragonal distortion in the films. The lattice 
parameters  are $a=b=5.836$~{\AA} and $c=5.852$~{\AA}~\cite{KOM13}. The compound 
has 26.4 valence electrons in the primitive cell and should exhibit a 
ferrimagnetic character with a total magnetic moment of 2.4~$\mu_B$ according to 
the Slater--Pauling rule. The magnetic moment as measured at room temperature by 
a vibrating sample magnetometer amounts to 2.57~$\mu_B$ and is slightly higher 
than the expected Slater--Pauling value. The deviation between experimental and 
theoretical value may be caused by the uncertainty in the mass density of the 
film.

\begin{figure}[htb]
 \centering
  \includegraphics[width=9cm]{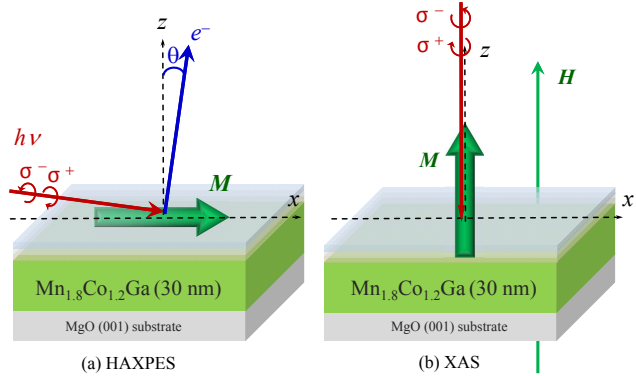}
  \caption{Sketch of the experimental geometry for a Mn$_{1.8}$Co$_{1.2}$Ga thin film:\\
     (a) HAXPES: remanently magnetised, \\
     (b) XAS: magnetised by external field $H$.}
\label{fig:setup}
\end{figure}

The HAXPES experiment with an excitation energy of 7.940~keV was performed at 
beamline BL47XU~\cite{OFF13,IKM13} of SPring-8. The energy distribution of the 
photoemitted electrons was analysed by using a hemispherical analyser 
(VG-Scienta R4000-12kV) with an overall energy resolution of 150 or 250~meV. The 
angle between the electron spectrometer and the photon propagation was fixed at 
$90^{\circ}$. The detection angle was set to $\theta=1^{\circ}$ to reach the 
near-normal emission geometry and to ensure that the polarisation vector of the 
circularly polarised photons is nearly parallel ($\sigma^-$) (antiparallel for 
$\sigma^+$) to the magnetisation $M$ (see Figure~\ref{fig:setup}a). The thin 
films were magnetised \textit{ex situ} along the direction of the photon beam 
(parallel to the surface plane) before being introduced into the UHV chamber. 
The polarisation of the incident photons was varied by using an in-vacuum phase 
retarder based on a 600-$\mu$m-thick diamond crystal with (220) 
orientation~\cite{SKM98}. The direct beam was linearly polarised with 
$P_p=0.99$. By using the phase retarder, the degree of circular polarisation was 
set to $P_{c} > 0.9$.

The XAS and XMCD investigations were performed with the magnetic circular 
dichroism measurement system at beamline BL25SU~\cite{NMG05} of SPring-8. The 
helicity switching of the circularly polarised radiation was performed by twin 
helical undulators. The absorption signal was measured in the total electron 
yield mode with the energy resolution set to $E/\Delta E = 5\times10^3$. The 
samples were magnetised in an induction field of $\mu_0H=\pm1.9$~T by a water-cooled-type 
electromagnet. A sketch of the experimental geometry for both 
measurements --~HAXPES and XAS~-- is shown in Figure~\ref{fig:setup}.

\section{Calculation details}
\label{sec:calcdetail}

The electronic structure calculations for the Mn$_{1.8}$Co$_{1.2}$Ga alloy with 
random site occupation were performed by means of the full-potential, fully 
relativistic spin-polarised Korringa--Kohn--Rostoker (SPR-KKR) 
method~\cite{EKM11} in combination with the coherent potential approximation 
(CPA)~\cite{Gyo72,But85}. The experimental lattice parameters were used because 
of the strain induced in the thin film. No remarkable differences in the results 
were observed when using the averaged cubic lattice parameter 
($\overline{a}=5.844$~\AA) or the structure with slight tetragonal distortion 
(see Section~\ref{sec:expdetail}). The exchange-correlation functional was taken 
within the generalised gradient approximation (GGA) in the parametrisation of 
Perdew--Burke--Enzerhof (PBE)~\cite{PBE96}. A $(22\times22\times22)$-based point 
mesh was used for integration in $k$ space; this mesh resulted in 1469 $k$ 
points in the irreducible wedge of the Brillouin zone. 

The non-integer 1.8~:~1.2~:~1 stoichiometry results in non-integer site 
occupations where some Co and Mn atoms are placed randomly in the $4d$ position of the regular 
lattice. This needs a special treatment in the calculations. The method used 
here to describe the random occupation of sites is the coherent potential 
approximation~\cite{Gyo72,But85}. The CPA can be easily performed within 
the Korringa--Kohn--Rostoker Green's function method. In the CPA, the random 
array of real on-site potentials is replaced by an ordered array of effective 
potentials and thus it describes the behaviour of an atom in a mean-field 
environment. The CPA is suited for any site occupation $0<x<1$ of Mn$_{2-x}$Co$_{1+x}$Ga.
Site resolved quantities --~that are partial density of states, values of the 
magnetic moments, or numbers of electrons~-- have been calculated in a 
Wigner--Seitz cell around each site. It was assumed that all Wigner--Seitz cells have 
the same size. The total magnetic moment and number of valence electrons were 
calculated for the complete primitive cell.

\begin{figure}[htb]
 \centering
  \includegraphics[width=7cm]{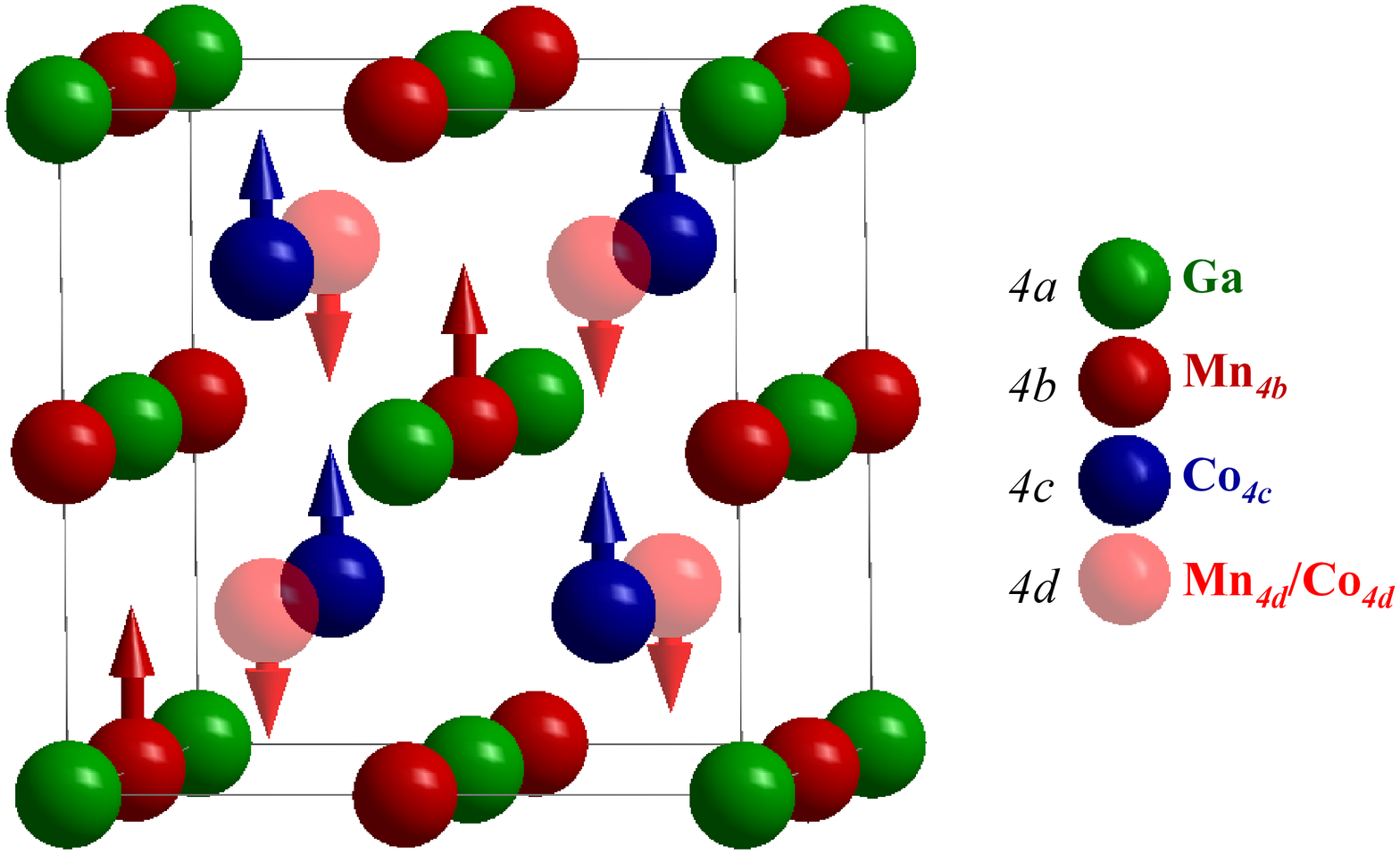}
  \caption{Cubic crystal structure of Mn$_{1.8}$Co$_{1.2}$Ga with space group {$F\bar{4}3m$}~(216).\\
          The Wyckoff positions $4a$, $4b$, and $4c$ are completely occupied by Ga, Mn, and Co, respectively.
           $4d$ contains a mixture of 80\% Mn and 20\% Co.
           It is seen that the centre Mn atom in the $4b$ position is octahedrally coordinated with respect to
           Ga in $4a$. The Co atom in $4c$ or the Mn and Co atoms in $4d$
           are tetrahedrally coordinated with respect to Ga. }
\label{fig:structure}
\end{figure}

The site occupation by the atoms of Mn$_{1.8}$Co$_{1.2}$Ga is shown in 
Figure~\ref{fig:structure}. The symmetry of the lattice belongs to space group 
$F\:\overline{4}3m$ (216). Ga occupies the Wyckoff position $4a$. Mn$_{4b}$ is 
placed at $4b$ and Co$_{4c}$ at $4c$. The Wyckoff position $4d$ is occupied 
randomly by 80\%  Mn$_{4d}$ and 20\% Co$_{4d}$. The directions of the local 
magnetic moments are indicated by arrows. 

Photoabsorption and XMCD spectra have been calculated using both many-electron 
and single-electron approaches. The single-particle calculations are based on 
the results of the electronic structure from the full-potential, fully 
relativistic spin-polarised KKR calculations. The core-levels themselves are 
strongly localized in the spherical part of the potential around the nuclei and 
behave like atomic states. Many electron effects are expected to appear in 
particular for the Mn atoms with localized electrons in the open $d$ shell. 
Therefore, atomic-type many-particle calculations were performed to explain some 
details of the Mn $2p$ states in the photon absorption. The multiplet 
calculations were performed using de~Groot's program {\sc ctm4xas}~\cite{SGr10}. 
This program includes also the effects of crystal fields and charge transfer. 
The details of the applied method are given in 
References~\cite{Cow81,Gro05,GKo08}. For the calculation of the $2p$ excitation, 
the Slater integrals were scaled to 90\% of their value from the Hartree--Fock 
calculations. The crystal field parameter was set to $10D_q=600$~meV and a 
magnetisation energy of $M=100$~meV was used for the calculation of the magnetic 
dichroism. The spectra were broadened by 200 to 600~meV according to the 
experimental resolution and lifetime broadening, with larger values used for the 
"$p_{1/2}$" parts of the spectra to account for Coster-Kronig contributions.

\section{Results and discussion}

\subsection{Electronic structure}
\label{sec:calc}

\begin{figure}[htb]
 \centering
  \includegraphics[width=11cm]{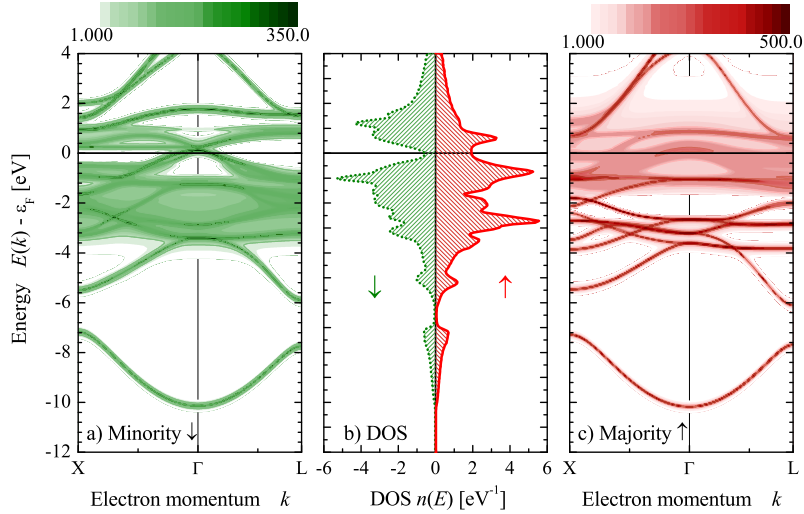}
  \caption{(Color online) Electronic structure of Mn$_{1.8}$Co$_{1.2}$Ga, showing the
       calculated minority (a) and majority (c) Bloch spectral function in the $\Delta$
       directions and the spin-resolved density of states (b).
       The grey (color online) scale is in atomic units.}
\label{fig:spagdos}
\end{figure}

Figure~\ref{fig:spagdos} shows the Bloch spectral functions for majority (a) and
minority (b) spin electrons together with the accompanying density of states
(DOS) (b). The minority channel exhibits a crossing point at $\Gamma$ close to
the Fermi energy $\epsilon_F$, which results in a low density of states. The
broadening of the majority band dispersion, as is seen from the Bloch spectral
function (Figure~\ref{fig:spagdos}c), is caused by the chemical disorder. This
broadening reduces the number of majority electrons at $\epsilon_F$. The
broadening results in a low effective spin polarisation at the Fermi energy.
This explains the rather low TMR values reported for the Mn$_{3-x}$Co$_{x}$Ga
system~\cite{KMM14}.

\begin{figure}[H]
 \centering
  \includegraphics[width=6cm]{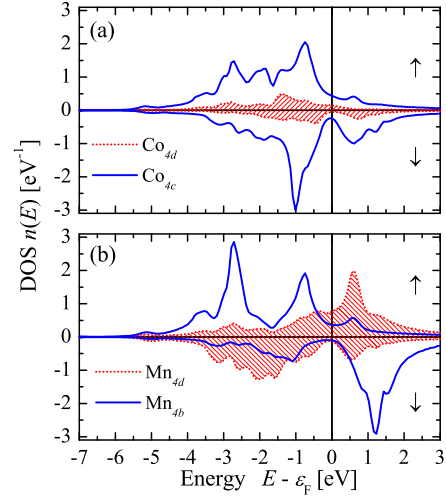}
  \caption{(Color online) Spin-resolved partial density of states
           of Co and Mn in Mn$_{1.8}$Co$_{1.2}$Ga.}
\label{fig:partdos}
\end{figure}

It is well known for the Heusler structure that octahedrally, main group element 
(here Ga) coordinated manganese atom --~in this case Mn$_{4b}$~-- exhibit highly 
localised $d$ electrons. To investigate the element-specific properties of Mn 
and Co in detail, the spin-resolved partial densities of states of Co and Mn at 
different sites in Mn$_{1.8}$Co$_{1.2}$Ga were calculated and are shown in 
Figure~\ref{fig:partdos}. 

The contribution of the Co$_{4d}$ states (shaded area in Figure~\ref{fig:partdos}a)
to the DOS is  weak because Co occupies only 20\% of the Wyckoff position
$4d$. The rather uniform distribution over the whole energy range of $d$
states indicates the delocalised character of the electrons in both spin
channels. The Mn$_{4d}$ states exhibit also a nearly uniform distribution over the
occupied valence bands in both spin channels. Only the unoccupied states of the
minority conduction band exhibit a clear maximum at 0.6~eV above $\epsilon_F$.
The minority valence band consists mainly of Co$_{4c}$ states, and the majority
states of Co$_{4c}$ are smoothly distributed over the entire valence band. The case
of the Mn$_{4b}$ states is different: They exhibit a pronounced localisation of the
valence electrons in the majority channel, resulting in sharp, peaked maxima at
$-0.74$ and $-2.7$~eV. At the same time, the localisation results in the peaked
DOS at 1.2~eV in the minority conducting band. The localised Mn states generate
a peculiar exchange splitting between the occupied and unoccupied states of up
to $\sim$4~eV. The Bloch spectral functions shown in Figure~\ref{fig:spagdos}
reveal that the localized Mn states are less affected by the chemical disorder
scattering. Using X-ray absorption or photoemission spectroscopy allows the described
behaviour of the valence $d$ states to be explored by investigating  the
interaction of $2p$ core holes and $3d$ valence electrons.

\begin{figure}[H]
 \centering
  \includegraphics[width=12cm]{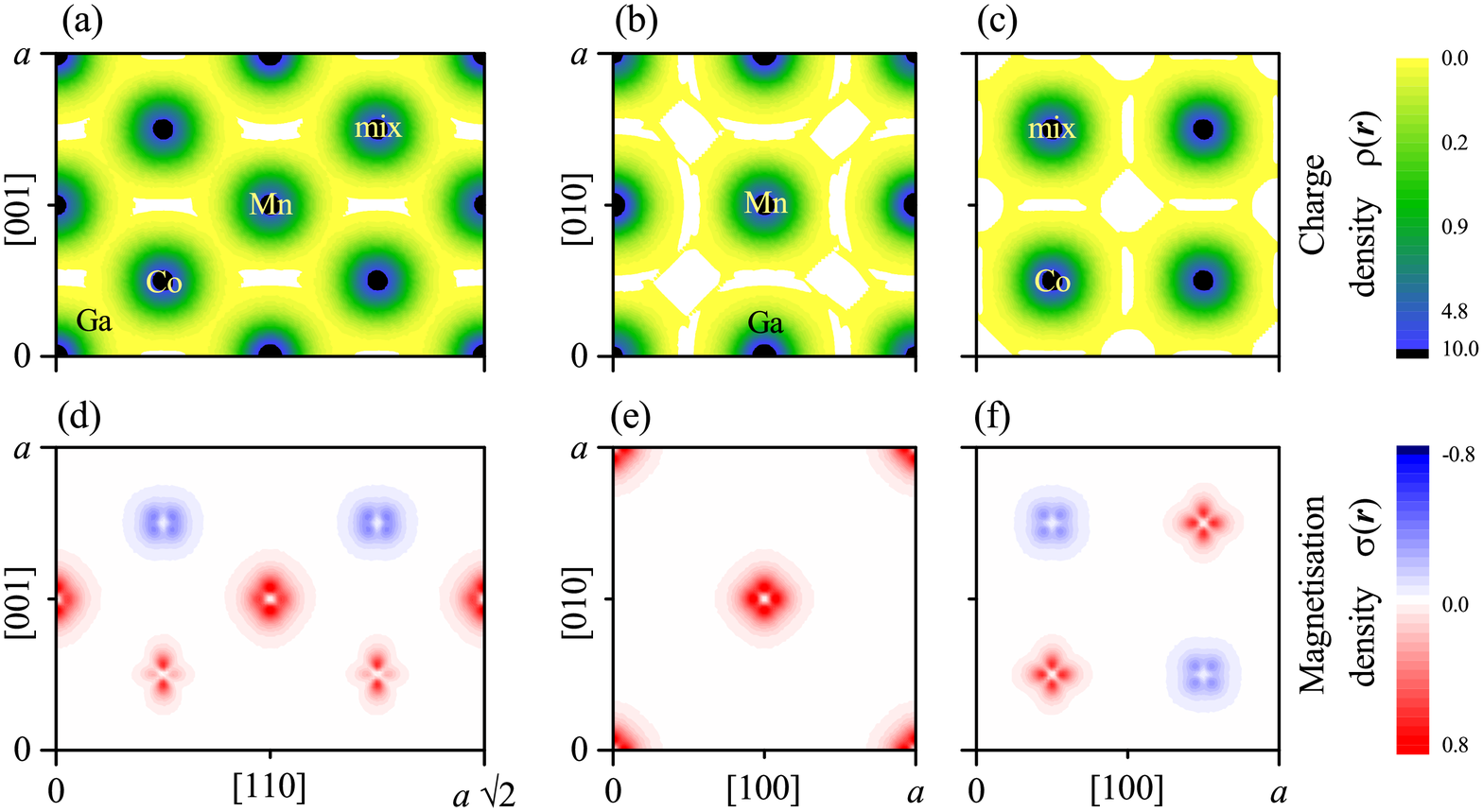}
  \caption{(Color online) Charge and magnetisation density of Co and Mn in Mn$_{1.8}$Co$_{1.2}$Ga.\\
            (a)...(c) are the charge densities ($\rho(r)$) and (d)...(f) the magnetisation densities ($\sigma(r)$).
            The values are given in atomic units as assigned by the colour (grey) scales.
            (a) and (d) are drawn in the (110) plane. The remaining densities are drawn in $\left<001\right>$-type planes with different 
            origins: $0,0,\frac{1}{2}$ in (b), (e) and $0,0,\frac{1}{4}$ in (c), (f). }
\label{fig:charge}
\end{figure}

Figure~\ref{fig:charge} shows the real space distribution of the charge and 
magnetisation densities in different planes of the cubic $fcc$ structure. The 
values at Wyckoff position $4c$ are averaged over the Mn and Co contributions in 
the mean field sense. Therefore, they are assigned in the Figures by {\it mix}. 
The charge densities ((a)...(c)) are total densities and calculated from the 
majority ($\rho_\uparrow$) and minority ($\rho_\downarrow$) spin densities 
including all core levels: $\rho(r)=\rho_\uparrow(r)+\rho_\downarrow(r)$. The 
magnetisation densities are calculated as difference of the spin densities: 
$\sigma(r)=\rho_\uparrow(r)-\rho_\downarrow(r)$.

\begin{figure}[H]
 \centering
  \includegraphics[width=12cm]{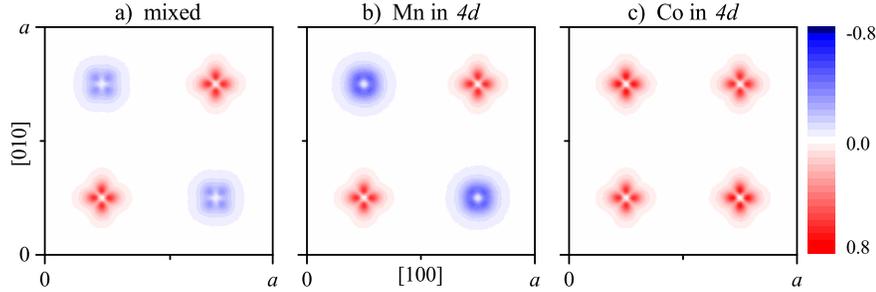}
  \caption{(Color online) Magnetisation density in the (001) plane with random
                          occupation of the $4d$ site by Co and Mn.\\
                          (a) shows the magnetisation density for 4d occupied by
                          80\% Mn and 20\% Co; (b) and (c) show $\sigma(r)$
                          for $4d$ occupied by 100\% Mn or Co, respectively.}
\label{fig:magndens}
\end{figure}

As mentioned above, the charge and magnetisation densities in the Wigner--Seitz 
cell around the $4d$ position are averaged over the occupation by Co and Mn 
atoms. In reality, however, there will be either a Co or a Mn atom in the $4d$ 
position but not both at once. Figure~\ref{fig:magndens} illustrates for the 
magnetisation density the situation where only one of the atom types occupies 
$4d$ and compares it to the average as was shown in 
Figure~\ref{fig:partdos}(f). As expected, $\sigma(r)$ at $4d$ is dominated by 
the part arising from the Mn atom that has its magnetic moment oriented with 
opposite sign to the total magnetic moment. The atom resolved magnetisation 
density reveals that the moment of Co in $4d$ is oriented parallel to the total 
magnetic moment. The magnetisation density of the Co atom in $4d$ has the same 
shape as that in $4c$. This behaviour is expected because such a occupation 
results in the regular Heusler structure with $F\:m\overline{3}m$ symmetry. 
Overall, the magnetisation densities of both types of Co atoms as well as of Mn 
in $4b$ are aligned along directions to the second nearest neighbours. Other 
than the Mn atom in $4b$ that has only transition metals as nearest neighbours, 
the Mn in $4d$ exhibits a magnetisation density that is aligned in the direction 
of the nearest neighbour atoms that are either Mn or Ga.

The calculated site-resolved spin and orbital moments per atom are listed in 
Table~\ref{tab:calcmag} together with the valence electron concentration and 
number of $d$ electrons at the different Mn and Co atoms. The site-resolved 
values were calculated for a Wigner--Seitz cell around the atoms. The total 
magnetic moment and number of valence electrons contain also the contribution 
from the Ga atoms. The total magnetic moment  $m_s+m_l$ is 
$m_{\rm calc}=2.44\:\mu_B$ in the primitive cell. This value agrees well with the 
measured saturation magnetic moment of $m_{\rm exp}=2.57\:\mu_B$~\cite{KOM13}.

\begin{table}[H]
\centering
  \caption{Calculated, total and element specific magnetic moments for Mn and Co in Mn$_{1.8}$Co$_{1.2}$Ga. \\
           The magnetic moments ($m$) are given in multiples of the Bohr magneton ($\mu_B$).
           $n_v$ and $n_d$ are the number of valence and $d$ electrons, respectively.
           Site specific values are per atom, total values are per primitive cell.}
     \begin{tabular}{llcccc}
     \br
                              &           & $m_s$  & $m_l$  & $n_v$ &  $n_d$ \\ 
     \mr
     \multirow{4}{*}{sites}   & Mn$_{4d}$ & -2.22  & -0.027 &  6.99 &  5.60 \\ \cline{2-6}
                              & Mn$_{4b}$ &  3.04  &  0.016 &  6.88 &  5.46 \\ \cline{2-6}  
                              & Co$_{4d}$ &  1.04  &  0.034 &  9.13 &  7.68 \\ \cline{2-6}
                              & Co$_{4c}$ &  0.98  &  0.040 &  9.27 &  7.75 \\
     \mr
     \multirow{2}{*}{average} & Mn        &  0.70  & -0.003 &  6.93 &  5.52 \\ \cline{2-6}
                              & Co        &  0.99  &  0.039 &  9.24 &  7.74 \\
     \mr
       {total}   & Mn$_{1.8}$Co$_{1.2}$Ga &  2.40  &  0.043 & 26.4  &        \\
     \br     
      \end{tabular}
\label{tab:calcmag}
\end{table}

\subsection{Core-level spectroscopy}
\label{sec:corelevel}

\subsubsection{Magnetic circular dichroism in angle-resolved photoelectron spectroscopy:}

The spectroscopy of $2p$ core levels is a powerful tool for studying the 
exchange interaction of the core holes with the valence electrons. The 
excitation of a core electron into a continuum state far above the Fermi level 
provides important information about the core--valence interactions. Especially, 
the MCD in photoemission enables an element-specific investigation of the 
magnetic properties of bulk materials~\cite{vdL00}, thin films, and 
intermetallic layers. The high bulk sensitivity of HAXPES in combination with 
circularly polarised radiation has been introduced to study element-specific 
properties of magnetic materials~\cite{KFS11}. For example, a pronounced MCD of 
Co $2p$ and Fe $2p$ states was reported for exchange-biased magnetic 
layers~\cite{KFS11} as well as for Co $2p$ and Mn $2p$ states in remanently 
magnetised half-metallic feromagnetic Co$_2$MnSi thin films~\cite{FEO14}. In the 
present work, MCD--HAXPES was used to investigate the magnetic properties of Mn 
and Co in Mn$_{1.8}$Co$_{1.2}$Ga thin films.

\begin{figure}[H]
  \centering
  \includegraphics[width=10cm]{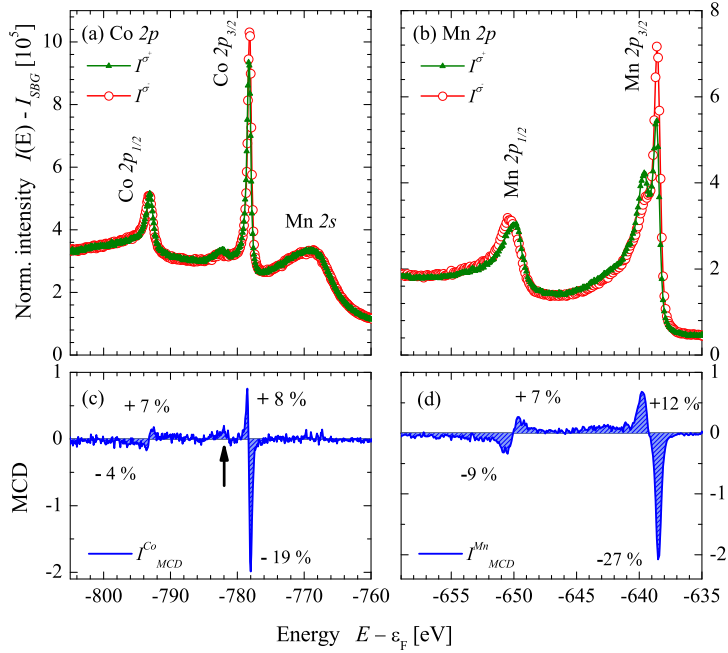}
  \caption{Polarisation-dependent photoelectron spectra of the Co $2p$ (a) and Mn $2p$ (b) core-level emission
           of Mn$_{1.8}$Co$_{1.2}$Ga and their difference (MCD) (c) and (d), respectively.
           Asymmetry values are marked at selected energies.
           Shown are the photoelectron spectra $I^+$ and $I^-$ and their
           difference $I_{MCD}$= $I^+ - I^-$ obtained with different helicity
           at fixed magnetisation parallel to the photon beam.}
  \label{fig:MCD}
 \end{figure}

Polarisation-dependent photoelectron spectra of the Co and Mn $2p$ core-level 
emission are shown in Figure~\ref{fig:MCD}. The MCD is characterised by an 
asymmetry that is defined as the ratio of the difference between the intensities 
$I^+$ and $I^-$ and their sum, $A=(I^+- I^-)/(I^++I^-)$, where $I^+$ corresponds 
to $\sigma^+$ and $I^-$ to $\sigma^-$ type helicity. The asymmetry values are 
calculated after subtracting a Shirley-type background from the spectra.

Figure~\ref{fig:MCD}(a) shows Co $2p$ and Mn $2s$ core-level spectra excited by 
7.938~keV, circularly polarised photons. The Mn $2s$ state at $-$769~eV did not 
exhibit a remarkable magnetic dichroism. The Co $2p$ states exhibit a spin-orbit 
splitting in $2p_{3/2}$ and $2p_{1/2}$ substates with $\Delta_{\rm SO}=14.9$~eV. 
The shape of the spectra is typical for metallic Co $2p$ states. This proves 
that the films are not oxidised. A satellite arising from oxidation should 
appear around 10~eV below the main $2p_{3/2}$ and $2p_{1/2}$ 
peaks~\cite{CYT05,GKo08}. A typical satellite often observed in materials with a 
face-centred cubic lattice is clearly resolved at 4.0~eV below the $2p_{3/2}$ 
maximum. Its intensity is enhanced by $\sigma^+$ excitation and reaches a 
positive asymmetry of 12\%. This peak may be due to interband transitions of the 
photoemitted electrons into an unoccupied Co state at 4~eV above $\epsilon_F$. 
The Co $2p$ states exhibit a clear dichroism that changes its sign in the 
sequence $- + + -$. The maximum asymmetry of the dichroism (19\%) is observed 
for the $2p_{3/2}$ state. The sequence of signs of the dichroism is typical for 
a Zeeman-type exchange splitting of the $2p$ states for their $m_j$ 
sublevels~\cite{Men98}.

A similar sign sequence of the dichroism is observed for the Mn $2p$ core 
states, as shown in Figure~\ref{fig:MCD}d. This confirms the parallel alignment 
of the magnetic moments of Mn and Co. The Mn $2p$ state exhibits an additional 
splitting, compared to the Co $2p$ state. The spin-orbit splitting of Mn $2p$ is 
$\Delta_{SO}\approx11.9$~eV. One may expect a splitting of the Mn $2p$ core 
states owing to a chemical shift because the Mn atoms occupy two different 
sublattices, Mn$_{4d}$ and Mn$_{4b}$. The calculated chemical shift, however, is 
only about 40~meV, but this value is clearly lower than the resolution of the 
experiment. Thus, it cannot explain the Mn $2p$ spectra. The Mn $2p$ core states 
did not show any hint on oxidation. It was shown previously, that in MnGa thin 
films, Ga-O bonds are formed at the interface, which prevent the oxidation of Mn 
at the interface~\cite{VOK14}. The additional splitting of the $2p_{3/2}$ state 
is attributed to the core hole--valence interaction. Such a splitting is caused 
by the Coulomb interaction of the $2p$ core hole and the $3d$ localised valence 
electrons of the majority channel. At the $2p_{3/2}$ state, this exchange 
splitting amounts to $\Delta_{\rm EX}=1.0$~eV, which is typical for Heusler 
alloys~\cite{YKS98,PSB99,OFB11}. The asymmetry values of the Mn states are 
higher than those of the Co states owing to the localised character of Mn $3d$ 
states.

\subsubsection{X-ray absorption spectroscopy:}

\begin{figure}[H]
  \centering
   \includegraphics[width=12cm]{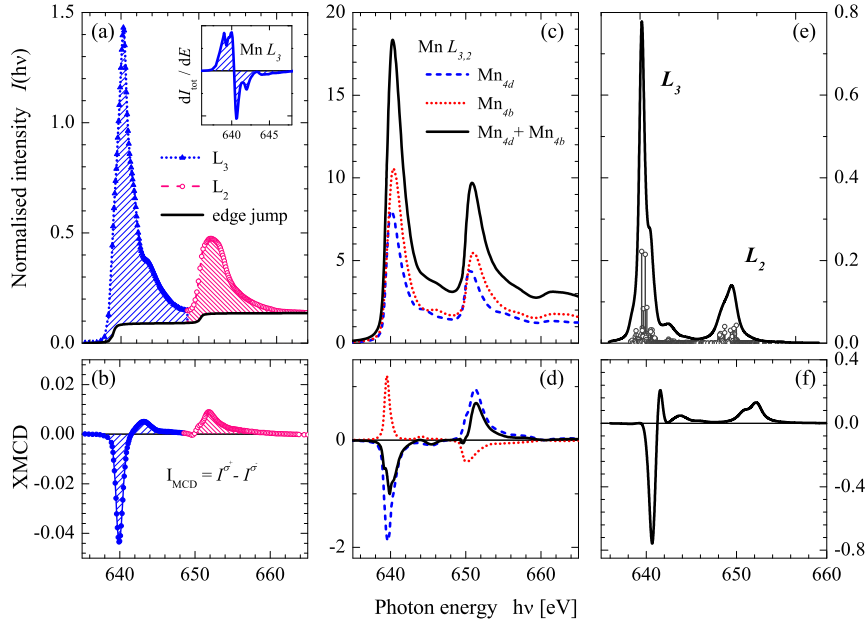}
   \caption{(Color online) Experimental (a) and calculated (c) XAS spectra at
   the Mn $L_{3,2}$ edges for Mn$_{1.8}$Co$_{1.2}$Ga. \\
   Corresponding XMCD spectra are shown in (b) and (d).
   (e) and (f) show the many electron calculated XAS spectra and corresponding 
   XMCD for an initial Mn$^{2+}$ state.
   Intensities of the experiment and many electron calculations are normalised to the number of $d$ holes,
   theoretical values from KKR calculations are in atomic units.}
  \label{fig:MnL32}
  \end{figure}

The MCD in combination with HAXPES has given information about the exchange 
effects of the core holes. The photoelectrons are detected at energies of 
unbound states far above the Fermi energy $\epsilon_F$ that have no direct 
relation to the part of the electronic structure responsible for transport 
properties. XAS and XMCD of Co and Mn $L_{3,2}$ edges were investigated and 
discussed in detail to investigate the exchange effect and to gain information 
about the unoccupied states close to $\epsilon_F$. The XAS spectra were 
normalised to the calculated number of unoccupied $d$ states for better 
comparison. The details of the data analysis are described in 
Reference~\cite{FEO14}. The experimental (XAS) and circular dichroism (XMCD) 
spectra of the Mn $L_{3,2}$ edges are shown in Figures~\ref{fig:MnL32}a and 
\ref{fig:MnL32}b. They are compared to calculated spectra in single-particle 
mode in Figures~\ref{fig:MnL32}c and \ref{fig:MnL32}d and atomic-type 
many-particle mode in Figures~\ref{fig:MnL32}e and \ref{fig:MnL32}f.

The spectra show the two $L_{3,2}$ white lines, corresponding to the  
$2p_{3/2} \rightarrow 3d$ and $2p_{1/2} \rightarrow 3d$ transitions, and reveal a high 
XMCD asymmetry with opposite sign. The corresponding single particle calculated 
XAS and XMCD spectra are shown in Figure~\ref{fig:MnL32}c and~\ref{fig:MnL32}d, 
respectively. The $2p_{3/2}$ single particle ground state binding energies of 
the two Mn and Co atoms in different positions differ by 100~meV and 10~meV, 
respectively. This is less than the resolution of the experiment and therefore 
not detectable. It merely results in a broadening of the lines.

From the calculations, Mn$_{1.8}$Co$_{1.2}$Ga is found to be a ferrimagnet where 
the two Mn atoms, Mn$_{4d}$ and Mn$_{4b}$, occupy two different sites with 
antiparallel spin orientation and different degrees of spin localisation (see 
Section~\ref{sec:calc}). The ferrimagnetic character of Mn in 
Mn$_{2.2}$Co$_{0.8}$Ga bulk material~\cite{KBA11} as well as in Mn$_2$CoGa thin 
films~\cite{MSK11} was previously confirmed by XAS--XMCD experiments.

At the $L_2$ edge a small step appears in the measured XMCD signal. The first 
derivative of the sum spectra is shown as an inset in Figure~\ref{fig:MnL32}a. 
Two maxima are clearly resolved, which indicates the superposition of two 
absorption maxima with slightly different energies. Similar to the observation 
in the electron emission spectra, the dichroism does not vanish between the 
$L_3$ and $L_2$ lines hinting on a mixing of the states. This mixing process is 
induced by the $p–-d$ Coulomb interaction. The influence of a $jj$-type mixing 
on the XAS spectra has been discussed in detail by Goering~\cite{Goe05}. The 
calculated many electron XAS spectra for the Mn$^{2+}$ $^6S_{5/2}$ initial state 
(see Figure~\ref{fig:MnL32}e) show a similar shape as the measured ones. The 
satellite appears at about 3.5~eV above the maximum of the $L_3$ line. The 
calculated splitting of the $L_2$ line is similar to that observed in the 
experiment. Both effects are attributed to multiplet effects and are similar to 
the behaviour of the Mn states of other Heusler compounds. More details about 
the multiplet effects are discussed in References~\cite{OFB11,FEO14}.

\begin{figure}[H]
  \centering
   \includegraphics[width=10cm]{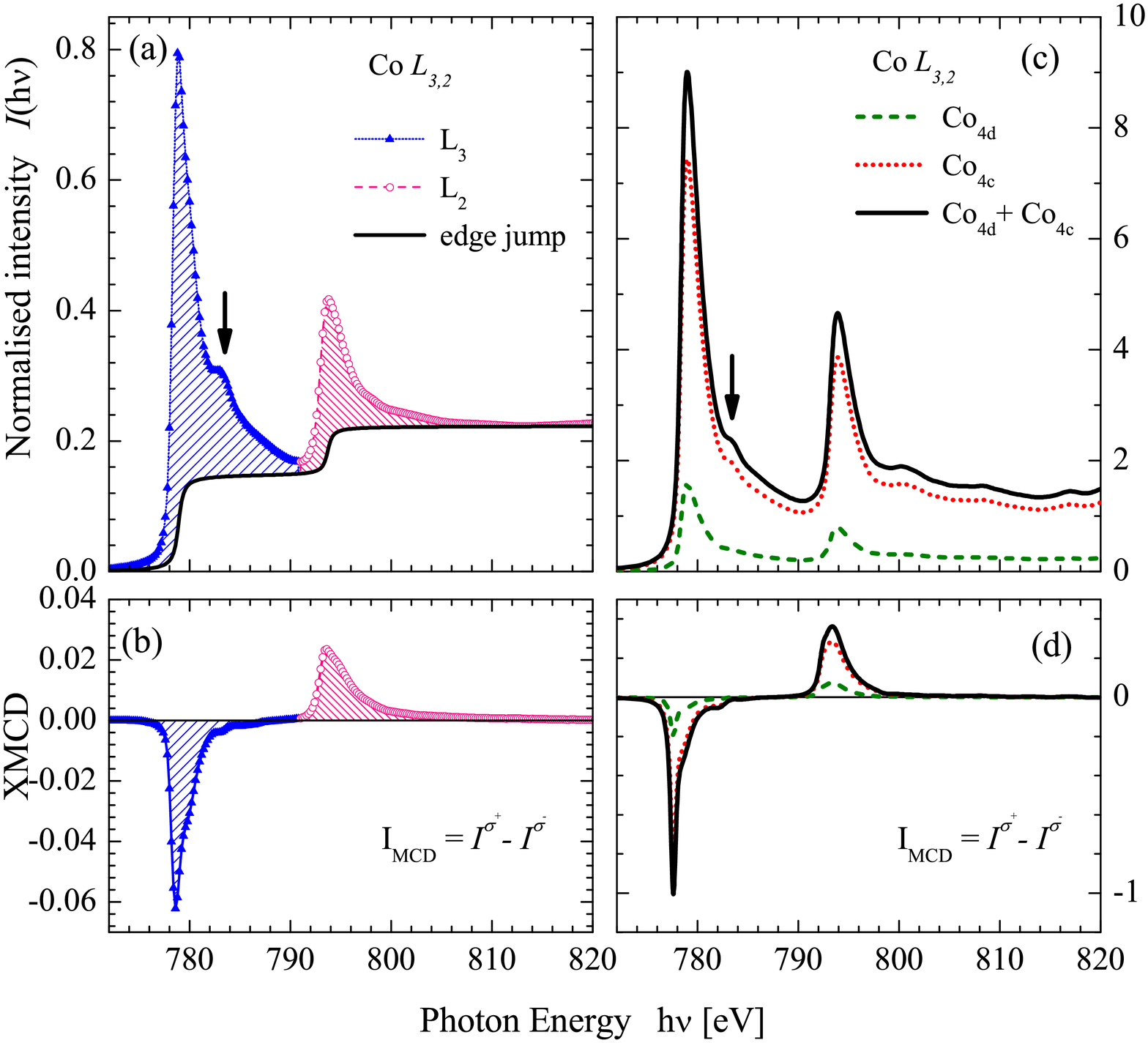}
   \caption{(Color online) Experimental (a) and calculated (c) XAS spectra
            of the Co $L_{3,2}$ edges for Mn$_{1.8}$Co$_{1.2}$Ga.
            The corresponding XMCD spectra are shown in (b) and (d), respectively.
            Intensities of the experiment are normalised to the number of $d$ holes,
            theoretical values are in atomic units.}
  \label{fig:CoL32}
  \end{figure}

The Co XAS and XMCD spectra are shown in Figure~\ref{fig:CoL32}. The sign order 
of the XMCD spectra confirms the parallel alignment of the Co magnetic moment to 
the mean of the Mn moment. Both calculated and measured Co $L_{2,3}$ XAS spectra 
show, in addition to the spin-orbit splitting of 14.8~eV, a shoulder at 4~eV 
above the $L_3$ white line (indicated by arrows). This is similar to the 4~eV 
satellite that is observed in the photoelectron spectrum of the $2p$ core level. 
This shoulder corresponds to a Co--Mn $sd$-band hybridisation state and was 
previously described as characteristic for highly ordered Heusler 
compounds~\cite{KEB06,KKS09}.

A sum-rule analysis was performed to obtain the spin and orbital magnetic 
moments from the XMCD data~\cite{Sto95,KKB09}. Three parameters enter the sum 
rule analysis that are the effective degree of circular polarisation  
($P_c \cos(\theta)$), the number of $d$-holes ($N_h$) and the spin correction factor 
($X$) such that the spin magnetic moment is given by:

\begin{equation}
	m_s = m_s^0 \frac{N_h}{X \: P_c \cos(\theta)},
\end{equation}

where $m_s^0$ is the uncorrected value for 100\% photon polarisation.  
$P_c \cos(\theta)$ is the effective polarisation projected on the direction of 
magnetisation.  $\theta$ is the angle between the photon beam and the direction 
of the applied magnetic field defining the magnetisation. In the present 
experiment, $P_c=0.96$ and $\theta=10^\circ$ resulting in an effective circular 
polarisation of 94.5\%. The number of unoccupied $3d$ states was assumed to be 
$N_h({\rm Mn})=4.5$, and $N_h({\rm Co})=2.3$ in accordance to the calculation of 
the occupied $d$ states. The values for the magnetic moments need to be 
corrected to account for the partial overlap of the $L_3$ and $L_2$ edges in 
Mn~\cite{TTJ96}. Here, the spin magnetic moment obtained by the sum rule 
analysis is corrected by the factors $X=0.68$ for Mn$^{2+}$ and 0.874 for 
Co$^{3+}$ as reported by Teramura {\it et al}~\cite{TTJ96}. Similar corrections 
were used by Chen et al. for Co (0.956)~\cite{CIL95} and in our previous 
work~\cite{FEO14}. The resulting magnetic moments, together with the calculated 
values are summarized in Table~\ref{tab:specmag}. The uncorrected average spin 
moment per atom is 0.468~$\mu_B$ for Mn, this value is in well agreement with 
the reported value of 0.47~$\mu_B$ in Mn$_{2}$CoGa thin films~\cite{MSK11}. The 
uncorrected average spin moment of 0.89~$\mu_B$ for Co derived from XMCD agrees 
well with the calculated value of 0.99~$\mu_B$. $m_l$ is small for both, Co and 
Mn, as expected because the orbital moments in cubic structures are quenched. 
The total moment from XMCD is in well agreement with the moment calculated 
(2.44~$\mu_B$) or measured by VSM (2.57~$\mu_B$).

\begin{table}[htb]
\centering
  \caption{Comparison of elements specific properties for Mn and Co in Mn$_{1.8}$Co$_{1.2}$Ga 
           derived from sum-rule analysis to calculated values. Uncorrected measured values from the sum rules
           including the effective polarisation are assigned by an asterix ($^*$).
           Core holes were assumed to be $N_h({\rm Mn})=4.5$, and $N_h({\rm Co})=2.3$.
           $X_{\rm Mn}=0.68$ and $X_{\rm Co}=0.874$ were assumed for the
           corrected values. }
     \begin{tabular}{lcccc}
     \br
                                 &        & m$_s$ & m$_l$  & m$_{tot}$             \\ 
     \mr
     \multirow{4}{*}{XMCD}       & Mn$^*$ & 0.468 &  0.005 & \multirow{2}{*}{1.94} \\
                                 & Co$^*$ & 0.888 &  0.022 &                       \\ \cline{2-5}
                                 & Mn     & 0.688 &  0.007 & \multirow{2}{*}{2.46} \\
                                 & Co     & 1.016 &  0.025 &                       \\
     \mr
     \multirow{2}{*}{calculated} & Mn     & 0.70  & -0.003 & \multirow{2}{*}{2.44} \\
                                 & Co     & 0.99  &  0.04  &                       \\
     \br
     \end{tabular}
\label{tab:specmag}
\end{table}

\subsection{Spin-resolved unoccupied density of states}

\begin{figure}[H]
  \centering
   \includegraphics[width=9cm]{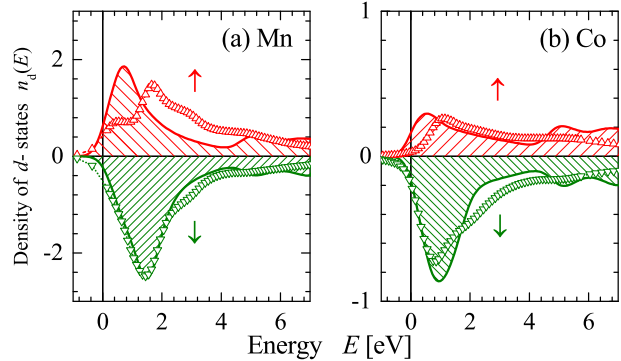}
   \caption{(Color online) Spin-resolved partial density of states extracted from the
             XAS--XMCD data measured at the $L_{3}$ edges of Mn$_{1.8}$Co$_{1.2}$Ga.
             Theoretical PDOS calculations are indicated by patterned areas.
             Majority and minority PDOS are shown on positive and negative scales, respectively.
             The calculated, partial densities of states are convoluted
             by a Fermi-Dirac distribution for 300~K and a Gaussian with 
             a width of 300~meV to account for the experimental broadening.}
  \label{fig:PDOS}
  \end{figure}

The $L_{3,2}$-XAS spectra are related to the unoccupied $d$-states. 
Figure~\ref{fig:PDOS} shows the spin-resolved unoccupied Mn and Co partial 
densities of states  derived from the $L_3$ edge XAS--XMCD data using the 
spin-resolved unoccupied PDOS function~\cite{KKB09,KKS09}. The shown calculated 
densities of states are convoluted by a Fermi-Dirac distribution for the 
unoccupied states at 300~K and by a Gaussian with a width of 300~meV to simulate 
the experimental broadening.

The majority and minority partial densities of states shown in 
Figure~\ref{fig:PDOS} are well resolved and clearly exhibit deviations from the 
calculated ones. The observed maxima of the minority densities agree well with 
the calculations for both, Mn and Co. The Co majority PDOS exhibits a shift of 
0.5~eV with respect to the maximum of the calculated majority states above 
$\epsilon_F$. The minority PDOS of Mn shows a pronounced maximum at 1.4~eV, 
which is dominated by unoccupied states located at Mn$_{4b}$ atoms (compare 
Figure~\ref{fig:partdos}). The majority PDOS of Mn shows a characteristic 
double-step increase and the observed maximum exhibits an energy shift of 
$\delta E=0.9$~eV. The corresponding maximum of the experimental majority PDOS 
of Mn appears at 1.65~eV. Besides correlation effects that will be explained 
below, this maximum may be effected by the multiplet effects or it may be due to 
spin-flip transitions~\cite{SAR10} into the high density of minority states that 
appears at the same energy.

The shifts of the maxima observed for the unoccupied minority PDOSs are 
characteristic of Heusler alloys~\cite{KKS09}. They are induced by electron 
correlation effects between localised and itinerant minority states. In 
particular for Co, the shift is a consequence of the itinerant bands that 
dominate the unoccupied majority Co states at the Fermi energy (see 
Figures~\ref{fig:spagdos} and~\ref{fig:partdos}). The transition energy for 
interaction of the core hole in the final state with localized $3d$ states is 
lower compared to the itinerant states because the latter screen the core hole 
to some extent~\cite{Bia82,WCA90}. This effect produces an energy shift between 
itinerant and localized states. The correlation energy of $\Delta E_c=0.5$~eV 
suggested in Reference~\cite{KKS09} agrees well with the shift observed in the 
present work. For Mn, an energy shift of $\delta E \approx 0.9$~eV is 
determined. This is in the same order as the shift observed for other Heusler 
alloys~\cite{KKB09,KBK10,KBA11,FEO14}. It is also induced by electron 
correlation effects between core hole and localised or itinerant minority 
states. Indeed, those states also differ for the majority electrons. The 
correlation energy $\Delta E_c$ strongly changes the minority PDOS determined 
from the XMCD measurements. As a consequence, the correct energy dependence of 
the unoccupied minority states can not be unambiguously detected with the used 
method.

\subsection{HAXPES of the valence band}

The occupied part of the electronic structure can easily be investigated by 
photoelectron spectroscopy. The linear dichroism in the angular distribution 
(LDAD) of photoelectrons is the difference between photoelectron currents 
ejected at a definite angle $\theta_\kappa$ by the linearly polarised light of 
two mutually perpendicular polarisations~\cite{CS93}:

\begin{equation}
 I_\kappa^{E_1}(\theta_\kappa) = \sigma_\kappa \left[ 1 + \beta_\kappa (3\cos^2\theta_\kappa-1)/2 \right] ,
\end{equation}

where $\sigma_\kappa$ is the partial photoionisation cross section and
$\beta_\kappa$ is the angular asymmetry parameter of the initial state with
quantum number $\kappa$. For linearly polarised light, the emitted electrons are
parallel $p$ ($\theta_\kappa = 0^{\circ}$) or perpendicular $s$
($\theta_\kappa=90^{\circ}$) to the surface normal.

\begin{figure}[htb]
 \centering
  \includegraphics[width=6cm]{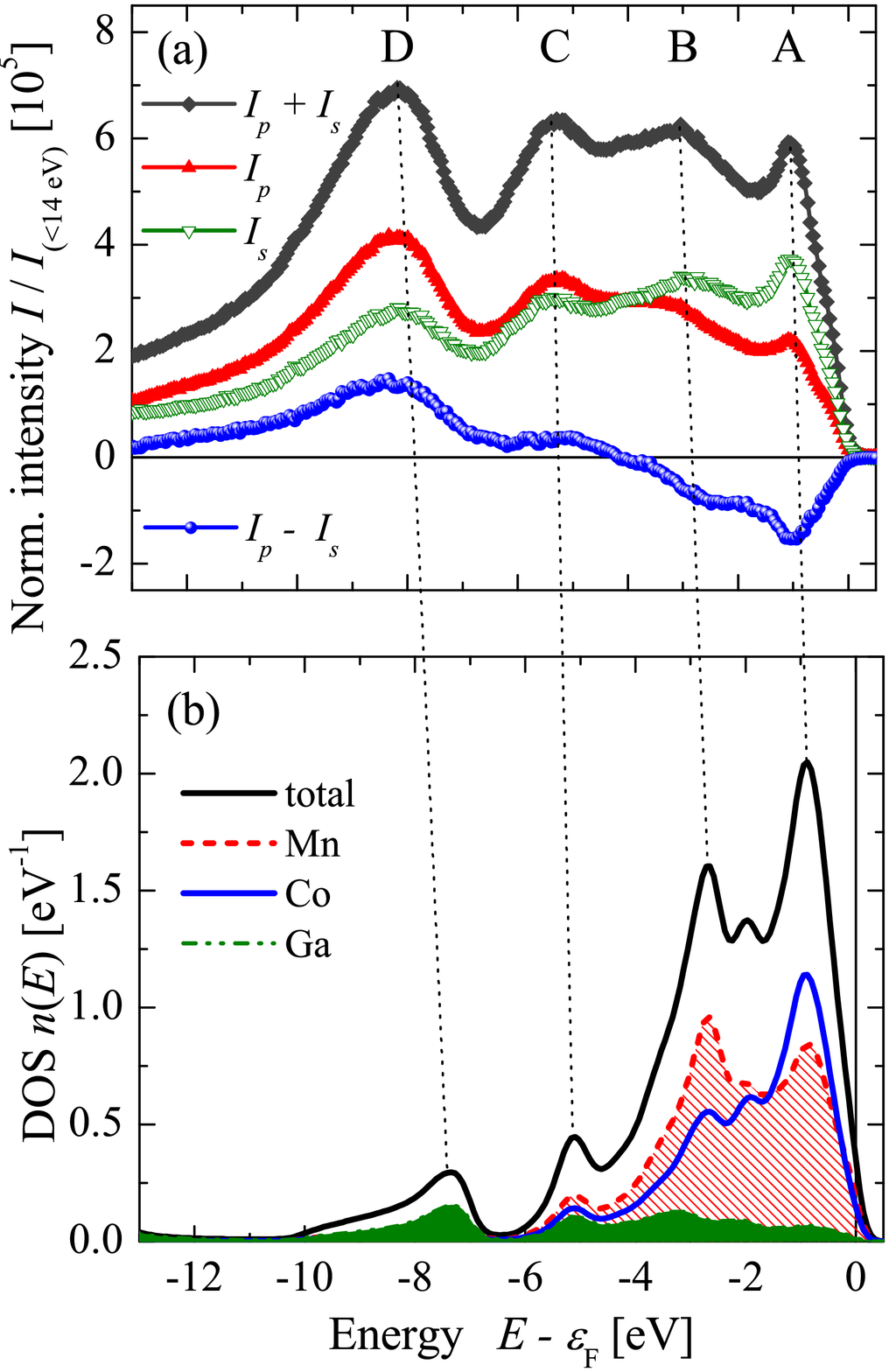}
  \caption{Valence band spectra (a) and total and partial density of states (b) of Mn$_{1.8}$Co$_{1.2}$Ga.\\
           The calculated, total and partial densities of states are convoluted
           by a Fermi-Dirac distribution for 300~K and broadened by a Gaussian with 
           a width of 150~meV corresponding to the experimental resolution.}
\label{fig:VB-LDAD}
\end{figure}

The polarisation-dependent photoelectron spectra and the calculated electronic 
structure of Mn$_{1.8}$Co$_{1.2}$Ga are compared in Figure~\ref{fig:VB-LDAD}.
The shown densities of states are convoluted
by a Fermi-Dirac distribution for the occupied states at 300~K 
and by a Gaussian with a width of 150~meV to simulate the experimental resolution.
The valence band spectra were normalised to the secondary background below the 
valence band ($\leq$$-$14~eV), as described in detail in~\cite{OFF13}. The 
difference $I_p-I_s$ shows the linear dichroism. The values given in 
Figure~\ref{fig:VB-LDAD}a are the calculated asymmetries at the most pronounced maxima 
of the spectra (labelled A to D).

The density of states exhibits three main maxima (A, B, and C)  in the energy 
range from -7~eV to $\epsilon_F$ as well as the split-off $s$ band (D) with 
$a_1$ symmetry. These structures are clearly observed in the photoelectron 
spectra (marked by dotted lines). One observes an energy shift and broadening 
when moving away from $\epsilon_F$. This is due to the lifetime effects of the 
photoexcited electrons during the photoemission process. The broadening is 
caused by increasing the imaginary part and the shift by the real part of the 
complex self-energy of the photoexcited electrons. The maximum at about -8~eV 
arises from the $s$ states located mainly at the Ga atoms. The characteristic 
Heusler $sp$ hybridisation gap at around -7~eV is clearly resolved. The 
maximum C at about -5.3~eV corresponds to an excitation of the $p$ states. The 
maximum B at -2.7~eV arises mainly from the Mn majority states, as seen from 
the spin-resolved DOS in Figure~\ref{fig:spagdos}b. The sharp state A at 
-0.95~eV comprises localised Mn$_{4b}$ states of the majority DOS. Such sharp 
states have been previously observed in the tetragonal Mn--Co--Ga 
system~\cite{OKF12}. Changing the Co amounts in Mn$_{3-x}$Co$_{x}$Ga  shifts 
this state closer to $\epsilon_F$, ending in a {\it band Jahn--Teller effect} 
that results in the tetragonal distortion of the crystalline 
structure~\cite{CKF12}. Changing the polarisation from $p$ to $s$ causes 
pronounce differences to appear in the spectra. $d$ states (A and B) with small 
binding energies have a higher intensity for $s$ polarisation. As in other cubic 
Heusler compounds~\cite{OFK11,KJO11,SOF12}, the intensity of the $a_1$ states is 
suppressed by $s$ polarisation, but it does not vanish, as expected for angular 
asymmetry parameter $\beta_0= 2$~\cite{TNY01}. This is due to the 
$sp$-hybridisation of the states in the cubic symmetry. The linear dichoism of the 
$p$ states is positive and small, reaching a value of +5\% (at -5~eV). The $d$ 
states (maximum A) at -0.94~eV exhibit the highest linear dichroism asymmetry 
of 29\% with negative values. Those states have an angular asymmetry parameter 
of $\beta_2<0$.

\section{Summary and conclusions}

In summary, the electronic and magnetic properties of Mn$_{1.8}$Co$_{1.2}$Ga 
thin films were investigated both theoretically and experimentally. 
Mn$_{1.8}$Co$_{1.2}$Ga is found to be a ferrimagnet where the two Mn atoms, 
Mn$_{4d}$ and Mn$_{4b}$, occupy two different sites with antiparallel spin 
orientation and different degrees of spin localisation. The electronic structure 
calculations of the alloy Mn$_{1.8}$Co$_{1.2}$Ga were performed by means of the 
fully relativistic spin-polarised Korringa--Kohn--Rostoker method. The chemical 
disorder in this alloy caused a broadening of the majority bands as shown in the 
Bloch spectral function, generating a reduction of majority electrons at the 
Fermi energy. The broadening results in a low effective spin polarisation at the 
Fermi energy. This explains the rather low TMR values reported for the 
Mn$_{3-x}$Co$_{x}$Ga system~\cite{KMM14}. The spin-resolved partial densities of state 
of Co and Mn at different sites revealed a nearly uniform distribution of 
Mn$_{4d}$ and Co states over the valence band range. The Mn$_{4b}$ atoms exhibit 
a pronounced localisation of valence electrons in the majority channel. The 
Bloch spectral functions reveal that the localized Mn states are less affected 
by the chemical disorder scattering. The calculated atom resolved magnetisation 
density reveals that the moment of Co in $4d$ is oriented parallel to the total 
magnetic moment and has the same shape as that in $4c$. The magnetisation 
density at $4d$ is dominated by the part arising from the Mn atom that has its 
magnetic moment oriented with opposite sign to the total magnetic moment. The 
valence states were further investigated by linear dichroism in the angular 
distribution. All states are well resolved and in agreement with the calculated 
electronic structure. The remarkably high linear dichroism in the valence band 
enables a symmetry analysis of the contributing states. Detailed magnetic 
properties were explored by excitation with circularly polarised X-rays. 
Element-specific magnetic moments and spin-resolved partial unoccupied densities 
of states were determined by using XAS and XMCD. A shifts of the maxima observed 
for the unoccupied minority states is induced by electron correlation effects 
between core hole and localised or itinerant minority states with different 
strength. XMCD was used to extract the site resolved magnetic moments. An 
experimental values of 2.46~$\mu_B$ agree well with the calculated magnetic 
moments of 2.44~$\mu_B$ or measured by VSM (2.57~$\mu_B$). Based on  
\textit{ab-initio} calculations in combination with the core-level spectroscopy, it was 
shown that one of the two Mn moments has a more localized character, whereas the 
other Mn moment and the Co moment are more itinerant. The similar sign sequence 
of the dichroism in photoelectron spectra for the Mn $2p$ and Co $2p$ core 
states confirms the parallel alignment of the magnetic moments of Mn and Co. The 
splitting of the Mn $2p_{3/2}$ state is caused by the Coulomb interaction of the 
$2p$ core hole and the $3d$ localised valence electrons of the majority channel. 

A possible way to improve the spin-transport in MnGa-system could be a a 
partially substitution of the main group element Ga, or a off-stoichiometric 
composition, similar to the regular Heusler compound Co$_2$MnSi that has shown 
the record tunnel magnetoresistance with a giant ratio of up to 1995\% at 4.2~K 
with variation of the Mn content~\cite{LHT12}. Beside the theoretical 
investigation, a extensive study of such system by spectroscopic methods --~as 
shown in the present work~-- will be a key to understand the properties of those 
alloys.

\bigskip
\ack

The authors gratefully acknowledge financial support by the DfG-JST (P~1.3--A and 
2.1--A in FOR 1464 ASPIMATT). S.~M. thanks JSPS Grants-in-Aid for Scientific 
Research (No. 25600070). The synchrotron based HAXPES and XMCD measurements were 
performed at BL47XU and BL25SU of SPring-8 with approval of JASRI, Proposal Nos. 
2012B0043, 2014A0043 and 2013A1909.


\bigskip

\bibliography{MnCoGa_TF}

\begin{thebibliography}{10}

\bibitem{MWS10}
S.~Mizukami, D.~Watanabe, E.~P. Sajitha, H.~Naganuma, M.~Oogane, Y.~Ando, and
  T.~Miyazaki.
\newblock {\em IEEE Trans. Magn.}, 56(6):1863, 2010.

\bibitem{MKW12}
S.~Mizukami, T.~Kubota, F.~Wu, X.~Zhang, T.~Miyazaki, H.~Naganuma, M.~Oogane,
  A.~Sakuma, and Y.~Ando.
\newblock {\em Phys. Rev. B}, 85(1):014416, 2012.

\bibitem{BFW07}
B.~Balke, H.~G. Fecher, J.~Winterlik, and C.~Felser.
\newblock {\em App. Phys. Lett.}, 90:152504, 2007.

\bibitem{WBF08}
J.~Winterlik, B.~Balke, G.~H. Fecher, C.~Felser, M.~C.~M. Alves, F.~Bernardi,
  and J.~Morais.
\newblock {\em Phys. Rev. B}, 77(5):054406, 2008.

\bibitem{WMW09}
F.~Wu, S.~Mizukami, D.~Watanabe, H.~Naganuma, M.~Oogane, Y.~Ando, and
  T.~Miyazaki.
\newblock {\em App. Phys. Lett.}, 94:122503, 2009.

\bibitem{KAM11}
T.~Kubota, M.~Araidai, S.~Mizukami, X.~Zhang, Q.~L. Ma, H.~Naganuma, M.~Oogane,
  Y.~Ando, M.~Tsukada, and T.~Miyazaki.
\newblock {\em App. Phys. Lett.}, 99:192509, 2011.

\bibitem{KMM12}
T.~Kubota, Q.~L. Ma, S.~Mizukami, X.~Zhang, Y.~Miura, H.~Naganuma, M.~Oogane,
  Y.~Ando, and T.~Miyazaki.
\newblock {\em Appl. Phys. Express}, 5:043002, 2012.

\bibitem{MWS11}
S.~Mizukami, F.~Wu, A.~Sakuma, J.~Walowski, D.~Watanabe, T.~Kubota, X.~Zhang,
  H.~Naganuma, M.~Oogane, Y.~Ando, and T.~Miyazaki.
\newblock {\em Phys. Rev. Lett.}, 106(11):117201, 2011.

\bibitem{MAS14}
Q.~Ma, A.~Sugihara, K.~Suzuki, X.~Zhang, S.~Mizukami, and T.~Miyazaki.
\newblock {\em Spin}, 4:1440024, 2014.

\bibitem{AWF11}
V.~Alijani, J.~Winterlik, H.~G. Fecher, and C.~Felser.
\newblock {\em App. Phys. Lett.}, 99:222510, 2011.

\bibitem{WCG12}
J.~Winterlik, S.~Chadov, A.~Gupta, V.~Alijani, T.~Gasi, K.~Filsinger, B.~Balke,
  G.~H. Fecher, C.~A. Jenkins, F.~Casper, J.~K{\"u}bler, G.-D. Liu, L.~Gao,
  S.~S.~P. Parkin, and C.~Felser.
\newblock {\em Adv. Mater.}, 2012.

\bibitem{OKF12}
S.~Ouardi, T.~Kubota, G.~H. Fecher, R.~Stinshoff, S.~Mizukami, T.~Miyazaki,
  E.~Ikenaga, and C.~Felser.
\newblock {\em Appl. Phys. Lett.}, 101(24):242406, 2012.

\bibitem{KOM13}
T.~Kubota, S.~Ouardi, S.~Mizukami, G.~H. Fecher, C.~Felser, Y.~Ando, and
  T.~Miyazaki.
\newblock {\em J. Appl. Phys.}, 113:17C723, 2013.

\bibitem{KMM14}
T.~Kubota, S.~Mizukami, Q.~L. Ma, H.~Naganuma, M.~Oogane, Y.~Ando, and
  T.~Miyazaki.
\newblock {\em J. Appl. Phys.}, 115(19):17C704, 2014.

\bibitem{LHT12}
H.-x. Liu, Y.~Honda, T.~Taira, K.-i. Matsuda, M.~Arita, T.~Uemura, and
  M.~Yamamoto.
\newblock {\em Appl. Phys. Lett.}, 101:132418, 2012.

\bibitem{CKF12}
S.~Chadov, J.~Kiss, and C.~Felser.
\newblock {\em Adv. Funct. Mater.}, 23:832, 2012.

\bibitem{Kob05}
K.~Kobayashi.
\newblock {\em Nucl. Instr. Meth. Phys. Res. A}, 547:98, 2005.

\bibitem{Kob09}
K.~Kobayashi.
\newblock {\em Nucl. Instr. Meth. Phys. Res. A}, 601:32, 2009.

\bibitem{FBO07}
G.~H. Fecher, B.~Balke, S.~Ouardi, C.~Felser, G.~Sch{\"o}nhense, E.~Ikenaga,
  J.-J. Kim, S.~Ueda, and K.~Kobayashi.
\newblock {\em J. Phys. D: Appl. Phys.}, 40:1576, 2007.

\bibitem{FBG08}
G.~H. Fecher, B.~Balke, A.~Gloskowskii, S.~Ouardi, C.~Felser, T.~Ishikawa,
  M.~Yamamoto, Y.~Yamashita, H.~Yoshikawa, S.~Ueda, and K.~Kobayashi.
\newblock {\em Appl. Phys. Lett.}, 92:193513, 2008.

\bibitem{OBG09}
S.~Ouardi, B.~Balke, A.~Gloskovskii, G.~H. Fecher, C.~Felser,
  G.~Sch{\"o}nhense, T.~Ishikawa, T.~Uemura, M.~Yamamoto, H.~Sukegawa, W.~Wang,
  K.~Inomata, Y.~Yamashita, H.~Yoshikawa, S.~Ueda, and K.~Kobayashi.
\newblock {\em J Phys. D}, 42:084010, 2009.

\bibitem{VOK14}
C.~E. ViolBarbosa, S.~Ouardi, T.~Kubota, S.~Mizukami, G.~H. Fecher,
  T.~Miyazaki, X.~Kozina, E.~Ikenaga, and C.~Felser.
\newblock {\em J. Appl. Phys.}, 116:034508, 2014.

\bibitem{KKO14}
X.~Kozina, J.~Karel, S.~Ouardi, S.~Chadov, G.~H. Fecher, C.~Felser,
  G.~Stryganyuk, B.~Balke, T.~Ishikawa, T.~Uemura, M.~Yamamoto, E.~Ikenaga,
  S.~Ueda, and K.~Kobayashi.
\newblock {\em Phys. Rev. B}, 89:125116, 2014.

\bibitem{SYH10}
A.~Sekiyama, J.~Yamaguchi, A.~Higashia, M.~Obara, H.~Sugiyama, M.~Y. Kimura,
  S.~Suga, S.~Imada, I.~A. Nekrasov, M.~Yabashi, K.~Tamasaku, and T.~Ishikawa.
\newblock {\em New Journal of Physics}, 12:043045, 2010.

\bibitem{KFS11}
X.~Kozina, G.~H. Fecher, G.~Stryganyuk, S.~Ouardi, B.~Balke, C.~Felser,
  G.~Sch{\"o}nhense, E.~Ikenaga, T.~Sugiyama, N.~Kawamura, M.~Suzuki, T.~Taira,
  T.~Uemura, M.~Yamamoto, H.~Sukegawa, W.~Wang, K.~Inomata, and K.~Kobayashi.
\newblock {\em Phys. Rev. B}, 84:054449, 2011.

\bibitem{OFK11}
S.~Ouardi, G.~H. Fecher, X.~Kozina, G.~Stryganyuk, B.~Balke, C.~Felser,
  E.~Ikenaga, and K.~Kobayashi.
\newblock {\em Phys. Rev. Lett.}, 107(21):036402, 2011.

\bibitem{SHI13}
A.~Sekiyama, A.~Higashiya, and S.~Imada.
\newblock {\em J. Electron Spectrosc. Relat. Phenom.}, 190:201–204, 2013.

\bibitem{TKL06}
N.~D. Telling, P.~S. Keatley, G.~van~der Laan, R.~J. Hicken, E.~Arenholz,
  Y.~Sakuraba, M.~Oogane, Y.~Ando, and T.~Miyazaki.
\newblock {\em Phys. Rev. B}, 74:224439, 2006.

\bibitem{SWW87}
G.~Sch{\"u}tz, W.~Wagner, W.~Wilhelm, P.~Kienle, R.~Zeller, R.~Frahm, and
  G.~Materlik.
\newblock {\em Phys. Rev. Lett.}, 58:737, 1987.

\bibitem{CSM90}
C.~T. Chen, F.~Sette, Y.~Ma, and S.~Modesti.
\newblock {\em Phys. Rev. B}, 42:7262, 1990.

\bibitem{OFF13}
S.~Ouardi, G.~H. Fecher, and C.~Felser.
\newblock {\em J. Electron Spectrosc. Relat. Phenom.}, 190:249, 2013.

\bibitem{IKM13}
E.~Ikenaga, M.~Kobata, H.~Matsuda, T.~Sugiyama, H.~Daimon, and K.~Kobayashi.
\newblock {\em J. Electron Spectrosc. Relat. Phenom.}, 190:180, 2013.

\bibitem{SKM98}
M.~Suzuki, N.~Kawamura, M.~Mizukami, A.~Urata, H.~Maruyama, S.~Goto, and
  T.~Ishikawa.
\newblock {\em Jpn. J. Appl. Phys.}, 37:L1488, 1998.

\bibitem{NMG05}
T.~Nakamura, T.~Muro, F.~Z. Guo, T.~Matsushita, T.~Wakita, T.~Hirono,
  Y.~Takeuchi, and K.~Kobayashi.
\newblock {\em J. Electron Spectrosc. and Relat. Phenom.}, 144-147:1035, 2005.

\bibitem{EKM11}
H.~Ebert, D.~K{\"o}dderitzsch, and Minar.
\newblock {\em Rep. Prog. Phys}, 74:096501, 2011.

\bibitem{Gyo72}
B.~Gyorffy.
\newblock {\em Phys. Rev. B}, 5:2382, 1972.

\bibitem{But85}
W.~H. Butler.
\newblock {\em Phys. Rev. B}, 31:3260, 1985.

\bibitem{PBE96}
J.~P. Perdew, K.~Burke, and M.~Ernzerhof.
\newblock {\em Phys. Rev. Lett.}, 77:3865, 1996.

\bibitem{SGr10}
E.~Stavitski and F.~M.~F. de~Groot.
\newblock {\em Micron}, 41:687, 2010.

\bibitem{Cow81}
R.~D. Cowan.
\newblock {\em {The Theory of Atomic Structure and Spectra}}.
\newblock California Press, Berkeley, 1981.

\bibitem{Gro05}
F.~M.~F. de~Groot.
\newblock {\em Coordin. Chem. Rev.}, 249:31, 2005.

\bibitem{GKo08}
F.~de~Groot and A.~Kotani.
\newblock {\em {Core Level Spectroscopy of Solids}}.
\newblock CRC Press, Taylor and Francis Group, Boca Raton, London, New York,
  2008.

\bibitem{vdL00}
G.~van~der Laan, S.~S. Dhesi, and E.~Dudzik.
\newblock {\em Phys. Rev. B}, 61:12277, 2000.

\bibitem{FEO14}
G.~H. Fecher, D.~Ebke, S.~Ouardi, S.~Agrestini, C.~Y. Kuo, N.~Hollmann, Z.~Hu,
  A.~Gloskovskii, F.~Yakhou, N.~B. Brookes, and C.~Felser.
\newblock {\em SPIN}, 4:1440017, 2014.

\bibitem{CYT05}
A.~Chainan, T.~Yokoya, Y.~Takata, K.~Tamasaku, M.~Taguchi, T.~Shimojima,
  N.~Kamakura, K.~Horiba, S.~Tsuda, S.~Shin, D.~Miwa, Y.~Nishino, T.~Ishikawa,
  M.~Yabashi, K.~Kobayashi, H.~Namatame, M.~Taniguchi, K.~Takada, T.~Sasaki,
  H.~Sakurai, and E.~Takayama-Muromachi.
\newblock {\em Nucl. Instr. Meth. Phys. Res. A}, 547:163–168, 2005.

\bibitem{Men98}
J.~G. Menchero.
\newblock {\em Phys. Rev. B}, 57:993, 1998.

\bibitem{YKS98}
Y.M. Yarmoshenko, M.I. Katsnelson, E.I. Shreder, E.Z. Kurmaev, A.~Slebarski,
  S.~Plogmann, T.~Schlath{\"o}lter, J.~Braun, and M.~Neumann.
\newblock {\em Eur. Phys. J. B}, 2(1):1, 1998.

\bibitem{PSB99}
S.~Plogmann, T.~Schlathoelter, J.~Braun, M.~Neumann, Y.~M. Yarmoshenko, M.~V.
  Yablonskikh, E.~I. Shreder, E.~Z. Kurmaev, A.~Wrona, and A.~Slebarski.
\newblock {\em Phys. Rev. B}, 60:6428, 1999.

\bibitem{OFB11}
S.~Ouardi, G.~H. Fecher, B.~Balke, A.~Beleanu, X.~Kozina, G.~Stryganyuk,
  C.~Felser, W.~Kloss, H.~Schrader, F.~Bernardi, J.~Morais, E.~Ikenaga,
  Y.~Yamashita, S.~Ueda, and K.~Kobayashi.
\newblock {\em Phys. Rev. B}, 84:155122, 2011.

\bibitem{KBA11}
P.~Klaer, B.~Balke, V.~Alijani, J.~Winterlik, G.~H. Fecher, C.~Felser, and
  H.~J. Elmers.
\newblock {\em Phys. Rev. B}, 84:144413, 2011.

\bibitem{MSK11}
M.~Meinert, J.-M. Schmalhorst, C.~Klewe, G.~Reiss, E.~Arenholz, T.~B{\"o}hnert,
  and K.~Nielsch.
\newblock {\em Phys. Rev. B}, 84(13):132405, 2011.

\bibitem{Goe05}
E.~Goering.
\newblock {\em Philosophical Magazine}, 85:2895, 2005.

\bibitem{KEB06}
M.~Kallmayer, H.~J. Elmers, B.~Balke, S.~Wurmehl, F.~Emmerling, G.~H. Fecher,
  and C.~Felser.
\newblock {\em J. Phys. D: Appl. Phys.}, 39:786 -- 792, 2006.

\bibitem{KKS09}
M.~Kallmayer, P.~Klaer, H.~Schneider, E.~A. Jorge, C.~Herbort, G.~Jakob,
  M.~Jourdan, and H.~J. Elmers.
\newblock {\em Phys. Rev. B}, 80:020406(R), 2009.

\bibitem{Sto95}
J.~St{\"o}hr.
\newblock {\em J. Electron Spectrosc. Relat. Phenom.}, 75:253, 1995.

\bibitem{KKB09}
P.~Klaer, M.~Kallmayer, C.~G.~F. Blum, T.~Graf, J.~Barth, B.~Balke, G.~H.
  Fecher, C.~Felser, and H.~J. Elmers.
\newblock {\em Phys. Rev.B}, 80:144405, 2009.

\bibitem{TTJ96}
Y.~Teramura, A.~Tanaka, and T.~Jo.
\newblock {\em J. Phys. Soc. Jpn.}, 65(4):1053, 1996.

\bibitem{CIL95}
C.~T. Chen, Y.~U. Idzerda, H.-J. Lin, N.~V. Smith, G.~Meigs, E.~Chaban, G.~H.
  Ho, E.~Pellegrin, and F.~Settle.
\newblock {\em Phys. Rev. Lett.}, 75:152, 1995.

\bibitem{SAR10}
D.~Steil, S.~Alebrand, T.~Roth, M.~Kraus, T.~Kubota, M.~Oogane, Y.~Ando, H.~C.
  Schneider, M.~Aeschlimann, , and M.~Cinchetti.
\newblock {\em Phys. Rev. Lett.}, 105:217202, 2010.

\bibitem{Bia82}
A.~Bianconi.
\newblock {\em Phys. Rev. B}, 26:2741, 1982.

\bibitem{WCA90}
P.~J.~W. Weijs, M.~T. Czyzyk, J.~F. van Acker, W.~Speier, J.~B. Goedkoop,
  H.~van Leuken, H.~J.~M. Hendrix, R.~A. de~Groot, G.~van~der Laan, K.~H.~J.
  Buschow, G.~Wiech, and J.~C. Fuggle.
\newblock {\em Phys. Rev. B}, 41:11899, 1990.

\bibitem{KBK10}
P.~Klaer, T.~Bos, M.~Kallmayer, C.~G.~F. Blum, T.~Graf, J.~Barth, B.~Balke,
  G.~H. Fecher, C.~Felser, and H.~J. Elmers.
\newblock {\em Phys. Rev. B}, 82:104410, 2010.

\bibitem{CS93}
N.~A. Cherepkov and G.~Sch{\"o}nhense.
\newblock {\em Europhys. Lett.}, 24:79, 1993.

\bibitem{KJO11}
X.~Kozina, T.~J{\"a}ger, S.~Ouardi, A.~Gloskovskii, G.~Stryganyuk, G.~Jakob,
  T.~Sugiyama, E.~Ikenaga, G.~H. Fecher, and C.~Felser.
\newblock {\em Appl. Phys. Lett.}, 99:221908, 2011.

\bibitem{SOF12}
R.~Shan, S.~Ouardi, G.~H. Fecher, L.~Gao, A.~Kellock, A.~Gloskovskii, C.~E.
  ViolBarbosa, E.~Ikenaga, C.~Felser, and S.~S.~P. Parkin.
\newblock {\em Appl. Phys. Lett.}, 101:212102, 2012.

\bibitem{TNY01}
M.~B. Trzhaskovoskaya, V.~I. Nefedov, and V.~G. Yarzhemsky.
\newblock {\em Atom Data Nucl. Data}, 77:97, 2001.

\end{thebibliography}
\bibliographystyle{unsrt}

\end{document}